# Spin-Orbit Interaction of Light Enabled by Negative Coupling in High Quality Factor Optical Metasurfaces

*Wenlong Gao[1], Basudeb Sain[1], Thomas Zentgraf[1]*

1. Paderborn University, Department of Physics, Warburger Str. 100, 33098 Paderborn, Germany.

**We study the negative couplings amid local resonances of photonic metasurfaces. In our analysis, we discover pseudo-spin-orbit coupled bulk modes leading to lines of circularly polarized radiation eigenstates in two-dimensional momentum space which were considered to exist only in three-dimensions. Our theoretical model is exemplified via a guided resonance dielectric metasurface that possesses Type-II Non-Hermitian diabolical points, from where the circular polarization lines emanate. The designed metasurface carries circular polarized radiation eigenstates in both the 0$^{th}$ and $\pm 1^{st}$ diffraction orders, allowing spin-selective light deflections. The high quality factor nature and field enhancement of the designed metasurface could lead to applications for spin-selective sensing, beam control and nonlinear optics. Our findings open a gateway for the design of near-field couplings assisted metasurfaces.**

Recent years have seen a plethora of research on the photonic polarization patterns in the momentum space, adding further knowledge to real-space singular optics [1]. Besides L(linear polarization) lines and C (circular polarization) points that have been extensively studied, bound states in the continuum and band degeneracies emerge, involving properties of energy bands. Among the singularities, C points bear significant interest owning to the accompanied observables including giant optical dichroism in transmission/reflectivity spectrum [2–6]. Circular polarizations are normally found at isolated points in 2-dimensional (2D) polarization pattern, due to the codimension 2 of its constrains [7]. Due to the same cause, C lines normally lives in 3D, and can only exist in 2D by special constructions, firstly discussed by J. F. Nye [8].

On the other hand, spin-orbit interaction related phenomena in optics has raised tremendous interests [9]. Substantial efforts have been dedicated to uni-directional guiding [10,11], photonic spin hall effect [12–16], unveiling some appealing fundamental physics of photons and their applications. Moreover, analogue of electrons' Rashba and Dresselhaus spin-orbit interactions with photons in liquid crystal [17], photonic graphene [18], and Pancharatnam Berry phase metasurface [19] reveal great prospect for photonic spin-orbit interactions empowered by these platforms. Among the photon manipulation platforms, subwavelength nature of metasurfaces could provide new degree of freedom, namely near field couplings that are explored in this work. Near field couplings, however, has only been scarcely studied in metasurfaces compared with properties stemming from local responses in meta-atoms. Existing studies include PT-symmetric metasurfaces, radiation mediated near field localizations and polarization-controls [20–22].

We hereby unveil the near-field couplings, specifically negative couplings (NC) as a powerful tool for manipulating radiation polarization states from subwavelength local resonance that can lead to spin-orbit coupled circularly polarized C lines in 2D. Meanwhile, NC are ubiquitous in condensed matter physics to take account for inserted magnetic flux quantum [23], and has been found use in various phenomena in topological states of matter, for instance, high order topological insulator [24], and non-abelian topological charges [25]. We present an effective Hamiltonian elucidating NC induced bulk modes with pseudo-spin-orbit couplings and the associated radiation polarizations. The effects are exemplified in a guided resonance silicon metasurface. We further find that the designed metasurface allows C lines in both the $0^{th}$ and $\pm 1^{st}$ diffraction orders, manifesting NC as compelling means for creating photonic devices with spin selective functionalities.

We start our analysis with introducing pseudo-spin-orbit interaction in a 1D tight binding model (TBM) with NC. Schematic of the TBM is depicted in Fig. 1(a). The real space Hamiltonian is $\mathcal{H} = \sum_n J q_{1,n} q_{2,n}^\dagger + \sum_n -J q_{2,n+1} q_{1,n}^\dagger + c.c.$ $q_{1(2),n}$ and $q_{1(2),n}^\dagger$ are creation and annihilation operators for the $n^{th}$ unit-cell's 1st (2nd) site. In the momentum space $\mathcal{H}(k) = 2J\sin(k)\sigma_y$, whose eigenstates are $[1; i]^T$ and $[1; -i]^T$. Dispersions of the modes are given in Fig. 1(c), and manifest pseudo-spin-orbit interaction since they correspond to the spinful photon states if the bases are mapped to orthogonal electromagnetic field components in the vacuum.

The TBM can be effectively constructed with $s$ and $p$ orbits. To elucidate, we examine the inner-cell coupling constants within the unit-cells $U_1$ and $U_2$ of the 1d model shown in Fig. 1(b). Note that $U_1$ can be transformed to $U_2$ by a $C_2$ rotation followed by a mirror operation $M$ solely on the $p$ orbit. Mathematically, $C_2$ and $M$ can be expressed by Pauli matrices $\sigma_x$ and $\sigma_z$ respectively. Therefore the inner-cell Hamiltonian expressed as $J\sigma_x$ within $U_1$ is transformed as $(C_2 M) J\sigma_x (C_2 M)^{-1} = -J\sigma_x$ to $U_2$, meaning the coupling constants are oppositely signed and up to a gauge choice. NC is thus realized with such construction. Note that it has been reported that adding an extra off-resonance site could realize NC as well [26].

Such configuration can be conveniently realized in coupled silicon waveguides (Fig.1 (d)) whose interactions can be well approximated by TBM due to the tight confinement of photons in the waveguides [27,28]. The designed metasurface involve two basic elements, namely, the single waveguide (SW) and the double waveguide (DW) coloured in grey and blue respectively. The inset of Fig. 1(d) depicts the top view of one unit cell of the Metasurface. Detailed geometrical parameters are summarized in Supplementary Table S1 [32]. .For simplicity, the waveguides are placed into a homogeneous environment, omitting the influence from higher diffraction orders in high refractive index substrates. Such design could be realized with free-standing technique [29], or utilizing ultralow refractive index materials, e.g. nano-pillar $SiO_2$ films as substrates [30].

Dispersions and the corresponding fundamental mode profiles are given in Fig. 2(a). DW holds spatially even and odd modes residing on lower and higher frequencies than respectively. The odd mode and SW's fundamental mode's dispersions are designed to intersect at around 230

THz (Fig. 2(a)), and are corresponding to the *p* and *s* orbits in the TBM (Fig. 1(b)). Numerical simulations [31] show that the waveguide array's *x*-direction dispersion at the intersection point is sine-functions like, resembling the TBM. Validity of the TBM is further confirmed by the simulated eigenmode profiles (Supplementary Information (SI) [32], Note 1). The 2D effective Hamiltonian near to the intersection point can be expressed as a type-II Dirac point [33]:

$$\mathcal{H}_{QQ} = T\delta k_y I + v_y \delta k_y \sigma_z + v_x \delta k_x \sigma_y \tag{1}$$

Here, $\delta k_{x,y}$ are the momenta in $x$ and $y$ directions deviating from the Dirac point located at momentum $(0, k_d)$, $v_{x,y}$ are the corresponding Fermi velocities, and $T$ is the tilting parameter that satisfies $T > v_y$. Due to time reversal symmetry, there are two DP at $k_d > 0$ and $k_d < 0$. In the following we focus on the DP in $k_d > 0$. Note that equation (1) has frequency as the eigen value, and there exists a direct transformation from it to the more commonly used momentum eigenvalue Hamiltonian for coupled waveguide systems [32].

To allow the couplings to ambient environment photon states, the waveguides are sinusoidally modulated as such that the mirror symmetry $M_x(x, y, z) \to (-x, y, z)$ centred at each waveguide remains, and are characterized by amplitudes $A_{1,2}$ and a relative shift $\phi$. Once the modulations of waveguides are turned on, equation (1) is mediated by the non-Hermicity effect due to leaky radiations to the continuum. Band dispersions before and after the modulation are illustrated in SI [32], Note 2. A quantum mechanical shell model was exploited to study the modifications to equation (1). Accordingly, the non-Hermitian effective Hamiltonian can be expressed as:

$$\mathcal{H}_{eff} = \mathcal{H}_{QQ} - \frac{i}{2}\Gamma\Gamma^\dagger + P.V. \int dE' \frac{\Gamma\Gamma^\dagger}{E-E'} \tag{2a}$$

$$|p\rangle = \Gamma^\dagger |q\rangle \tag{2b}$$

The guided wave resonance $|q\rangle$ and radiation channels $|p\rangle$ are contained within subspace $\{Q\}$ and $\{P\}$ respectively and $\Gamma^\dagger$ is the coupling matrix [34]. The Cauchy principal value (P.V.) term represents the collective Lamb frequency shift due to couplings to radiation continuum [35–38]. In plasmonic metasurface the P.V. term can be delicately designed to counteract $\mathcal{H}_{QQ}$, leaving the non-Hermitian term dominant [20]. In our case, the P.V. term is substantially smaller than $\mathcal{H}_{QQ}$ and thus only accounted for slight frequency shifts. The approximated form of $\Gamma^\dagger$ to the first order can be expressed as [32]

$$\Gamma^\dagger = \begin{bmatrix} \kappa & \eta e^{i(\gamma+\Delta)}\delta k_x/k_d \\ -\kappa e^{i\gamma}\delta k_x/k_d & \eta e^{i\Delta} \end{bmatrix} \quad (3)$$

$\kappa(A_1)$, $\eta(A_2)$ are coupling strengths determined by $A_{1,2}$. $\Delta$ is a relative phase delay controlled by $\Phi$, and $\gamma$ is a phenomenological phase delay taking account for the deformation of the waveguide modes according to $k_x$. Each term in $\Gamma$ can be retrieved from numerical simulations [33]. When it satisfies $\kappa = \eta$ and $\Delta = 0$, $\Gamma$ can be simplified as $\kappa I + i\kappa e^{i\gamma}\frac{\delta k_x}{k_d}\sigma_y$. In such case we have $\Gamma\Gamma^\dagger = \kappa^2(I - 2\frac{\delta k_x}{k_d}\sin(\gamma)\sigma_y)$ to the first order in $\delta k_x$. Further Cooperating changes in $\delta k_y$ the 2D effective Hamiltonian can thus be expressed as:

$$\mathcal{H}_{eff} = \mathcal{H}_{QQ} - i(\tfrac{1}{2}\kappa^2 + v_t\delta k_y)I + iv_s\delta k_y\sigma_z + i\frac{\kappa^2\sin(\gamma)}{k_d}\delta k_x\sigma_y \quad (4)$$

$v_{t,s}$ are the tilting and splitting parameters of the imaginary part along $\delta k_y$. Note that commutative relations $[\mathcal{H}_{QQ}, \Gamma\Gamma^\dagger]|_{\delta k_y=0}$ and $[\mathcal{H}_{QQ}, \Gamma]|_{\delta k_y=0}$ hold, thus guarantee the circular polarized eigenstates in $\{P\}$ on $\delta k_y = 0$. We give the dispersions and polarization eigenstates of the theoretical model in [32].

In simulations of the realistic structure, we apply Bloch-periodic boundary conditions in the x−y plane and scattering boundary conditions in the z direction, and solve for the complex-valued eigen-frequencies. We find $A_1 = 10$ nm and $A_2 = 14.4$ nm can satisfy $\kappa = \eta$, which is well within the reported fabrication capabilities of sinusoidal sidewall waveguides [39], while keeping the Q factor in interested regime high as 7000. Furthermore, it is found that $\Phi = 0.25$ rad to satisfy $\Delta = 0$ [32]. The numerically simulated band dispersions' real and imaginary parts are shown in Fig. 2(c, d). For better visualization, only their differences are shown. Total dispersions are demonstrated in SI [32], Fig. S1. Notice that the real part maintain the Dirac point's linear dispersions while the imaginary presents crossing line nodes (CL) intersecting at the band touching. To distinguish it from the unmodulated case, the band touching is referred to as diabolical point (DP) from here.

Polarizations of radiation eigenstates are obtained by integrating the electrical fields at a plane above the metasurface [40]. In Fig. 2(e, f) we show the polarizations' degree of circular polarization $\sigma_d$ whose definition and the relation to photonic spin vector $\vec{S}$ are given in SI [32], Note 3. C lines with $\sigma_d = \pm 1$ emerge. A schematic illustration of the polarization structure around the C lines is shown in Fig. 2 (b). The C lines are also accompanied by orthogonal L lines that intersect at the DP. These properties can be properly described by equation (4), showing excellent agreements with retrieved parameters from the numerical simulations [32]. Notice that when it satisfies $\kappa \neq \eta$, DP can be reduced to a pair of exceptional points (EP) while CL split into two open arcs emanating from the EP and an isolated arc, as is shown in Fig. 3(a). More importantly, the C-lines can be instantaneously lifted when $\Delta \neq 0$ [32].

To inspect fidelity of the C-lines in the numerical simulations, we project normalized Stokes vectors of the polarizations of radiation eigenstates on Poincaré sphere for various Φ (Fig. 3(b)). As the momentum evolved on a circle around the DP (grey dashed-line in Fig. 3(a)). The momentum loop's start and end points are on the $k_x = 0$ line, whose polarizations are linear ($S_{1,3} = 0$). Note that as Φ increases from -0.25 to 0.7, the polarization trajectory sweeps different sides of the north/south poles of the Poincaré sphere, meaning there exists at least one value of Φ such that the trajectory travels exactly through the north/south poles corresponding to circular polarization state. The winding on the Poincaré sphere is a manifestation of the nontrivial $\pi$ Berry phase of the Dirac cone in the subsystem $\{Q\}$ [41,42].

Numerically simulated dispersion diagram along $k_x$ at the diabolic point with the $\sigma_d$ as hues is shown in Fig. 4(a). In proximity to DP, the dispersions consist of two bands with $\sigma_+(\sigma_d = 1)$ and $\sigma_-(\sigma_d = -1)$, circular polarized radiation eigenstates. At larger $k_x$, an excessive mode interfere with the $\sigma_\pm$ bands exhibiting elliptical polarization. The $\sigma_\pm$ bands can be manifested in the transmission/reflectivity spectrum for $\sigma_\pm$ incident light. Fig. 4 (c, d) shows that under $\sigma_\pm$ incident light, only one corresponding $\sigma_\pm$ bands shows an appreciable reflectivity approaching one. Note that the $\sigma_\pm$ incident photon are mainly reflected at $\sigma_\mp$ bands, since $\sigma_d$ is defined according the wave's propagation direction $\vec{k}$ to be consistent with photon's spin $\vec{S}$ (SI [32], Note 1). In our case the incident photon mainly excites the resonance whose polarization eigenstates's $\vec{S}$ aligns.

The detailed reflectivity spectrum near to the marked frequency noted by the stars in Fig. 4(c) are given in Fig. 4(b). The spectrum resembles an antisymmetric shape due to background reflections of the bare waveguides [43]. Moreover, at $\sigma_\pm$ bands it exhibit a mini-resonance peak for $\sigma_\pm$ incident. This arises due to the noncollinearity between the incident photon's spin vectors $\vec{S}$ to that of the eigen-polarization. In fact, the mini-resonance peak disappears when the time reversal of radiation eigenstate is taken as the incident (SI [32], Note 3).

Previous reports on metasurface with optical activity for spin photons [2,19,44] rely on optical responses that was determined by either the chiral geometry of individual meta-atoms or their collective coherent interference. The mutual couplings, specifically, NC introduced spin-polarization control was still in non-existence. The method we propose here is general, promising in being extended to other platforms with strong local resonances like plasmonic metasurfaces [45] and photonic crystals defect states [46] without trial-and-error.

Even richer spin polarization phenomena can be discovered in the diffraction orders. In BIC structures, diffraction orders are deliberately omitted since they can be smeared out imposing extra radiation channels [47]. In our case, however, radiation states in diffraction orders are found to be spin-polarized as well, which can enable spin-selective optical diffraction. $\sigma_d$ for the $\pm 1$ diffraction orders are depicted as hues in the dispersion diagram in Fig. 5(a). An intriguing handedness flipping emerge, noticing $\sigma_d$ are opposite to the $0^{th}$ order (Fig. 4(a)).

The handedness-flipping between diffraction orders is a generic phenomena to our design. We demonstrate a geometrical proof in Fig. 5(b). The radiated spherical wave fronts are depicted in dashed and solid lines respectively for each waveguide. Now suppose in the $0^{th}$ order, the radiated beam is in constructive interference at angle $\theta_1$, while the only allowed diffraction order is at angle $\theta_2$. Let $k_0$ be the free space wavenumber, we have the equality: $cos\theta_1 - cos\theta_2 = \frac{2\pi}{ak_0}$. $\lambda$ denotes the wavelength, and $R_1 = L_1 = N\lambda$, where $N \in \mathbb{Z}^+$. Inspecting the geometry, we have $R_2 = N\lambda + \frac{a}{2}cos\theta_1$ and $L_2 = N\lambda + \frac{a}{2}cos\theta_2$, and consequently $R_2 - L_2 = \frac{\lambda}{2}$, meaning there exists a $\pi$ phase difference between the red and the blue wave fronts. This establishes that if the $0^{th}$ order radiation constructed with the red wave fronts is $\sigma_+$, the $1^{st}$ diffraction order with the blue wave fronts is $\sigma_-$. An alternative explanation of the spin-flipping is by noticing that the orbit centre of $|q_1\rangle$ (SW) and $|q_2\rangle$ (DW) are $a/2$ in distance. Then momentum space periodicity of the total wavefunction demands $u_{k_x+\Lambda}(x) = e^{i\Lambda x} u_{k_x}(x)$, where $u_{k_x}(x) = \langle x|q_1\rangle + i\langle x|q_2\rangle$ is the periodic part of Bloch wave function along $x$, and $\Lambda$ is reciprocal vector. Finally we have $u_{k_x+\Lambda}(x) = \langle x|q_1\rangle - i\langle x|q_2\rangle$.

In conclusion, we have studied the spin-orbit interactions of a high-Q metasurface with near field negative couplings. Lines of circularly polarized radiation eigenstates are discovered. The high-Q and spin selective feature allow strong light-matter interaction for a specific photonic spin for incident photons. Strikingly the metasurface possesses fully spin-polarized states even in the $\pm$1st diffraction orders whose handedness are opposite to the 0th order. This feature could allow spin-selective beam steering and diffractions.

**Acknowledgements**

This project has received funding from the European Research Council (ERC) under the European Union's Horizon 2020 research and innovation programme (grant agreement No 724306).

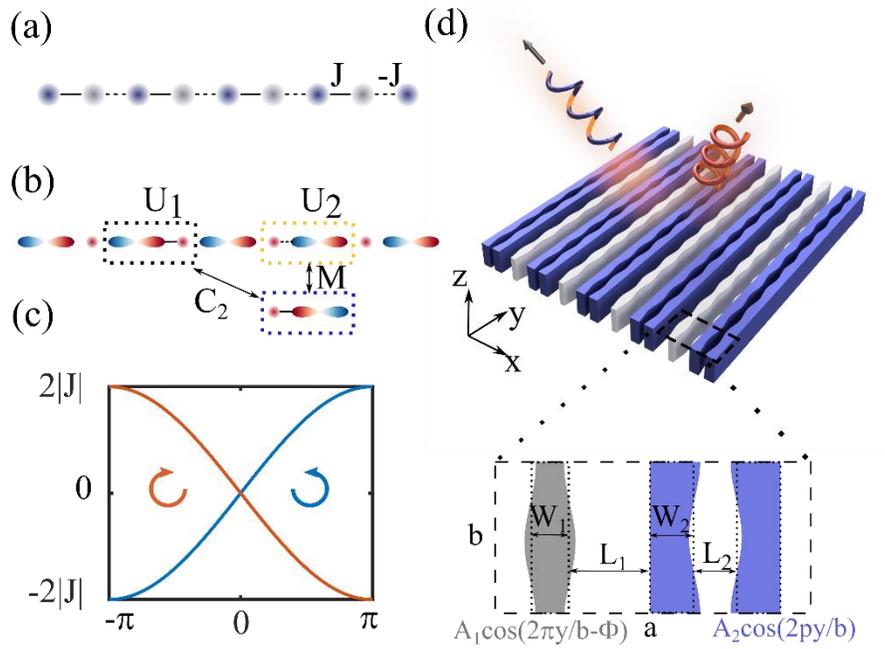

Figure.1 (a, b) Schematic of the tight-binding model with negative couplings and the construction with p and s orbits. (c) Dispersion of the bulk chiral modes of the tight binding model (d) Schematic of the silicon metasurface. The resonance modes in the metasurface are coupled to the circular polarizations states in the continuum with the sinusoidal modulations. The inset shows the detailed top view of the detailed geometrical parameters of the metasurface.

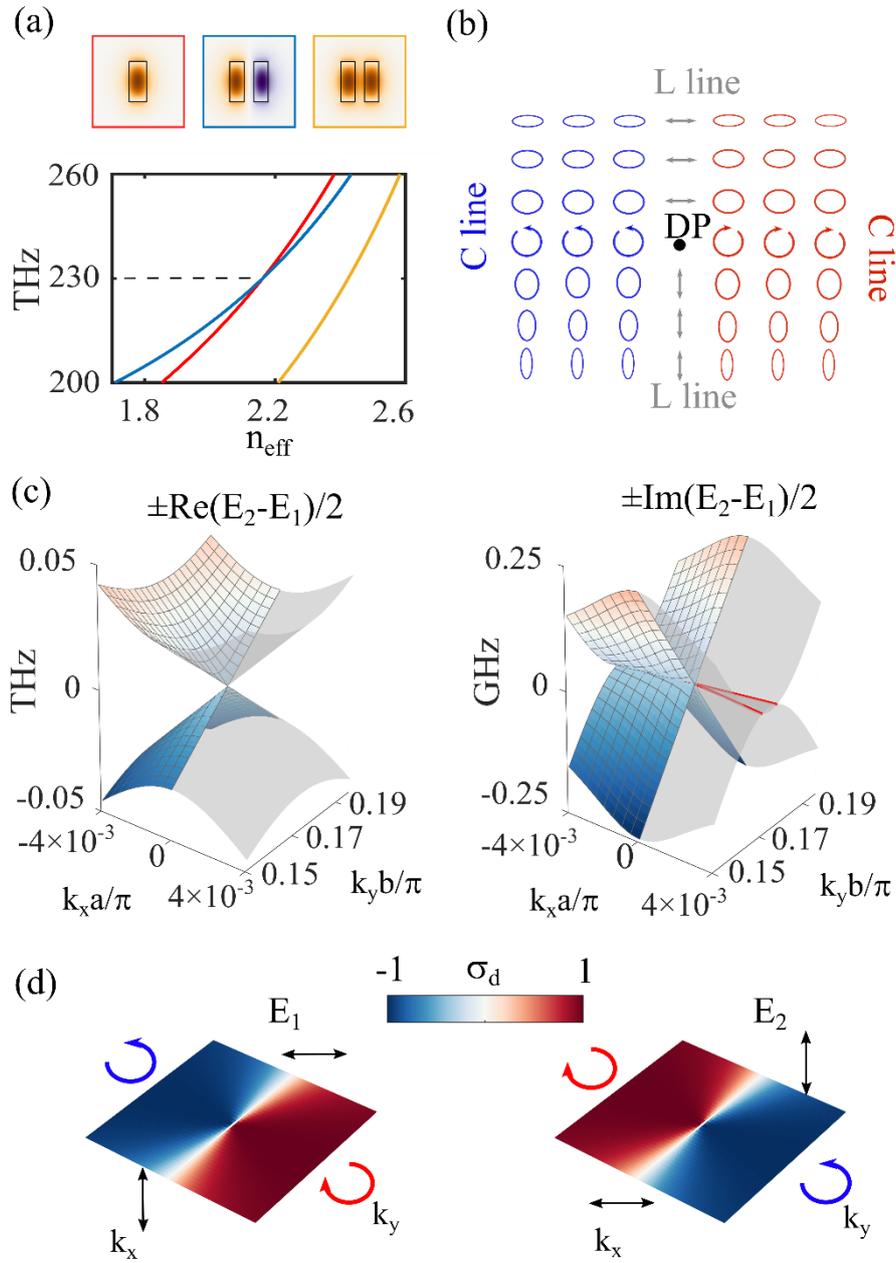

Figure. 2 (a) Dispersion of the individual waveguide modes without sinusoidal modulations. The mode profiles' $E_y$ components are given in the correspondingly coloured insets. (b) Schematic of the C lines that emanates from the diabolic points (DP) and are orthogonal to the L lines. (c) Real and imaginary part of the difference of eigen frequencies $E_1$ and $E_2$. The real part resembles a lossless Dirac cone while the imaginary part obtains two crossing line nodes noted by the solid red lines. (d) Eigen polarizations' $\sigma_d$ parameters of the two bands around the Diabolic cone.

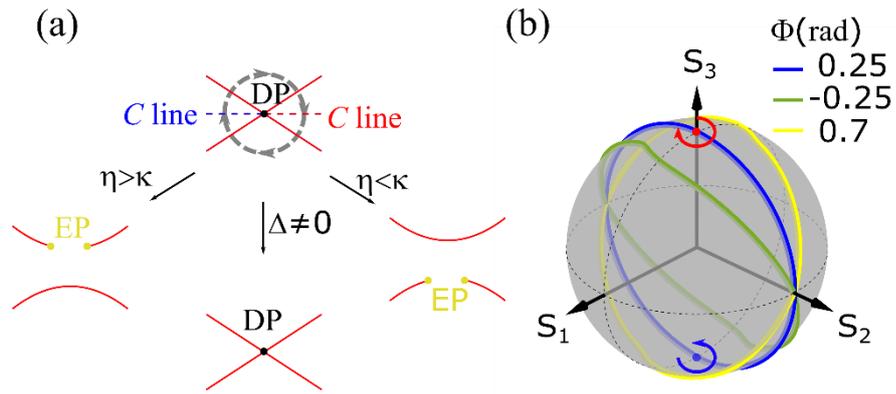

Figure. 3 (a) Illustration of the imaginary part's crossing line nodes (red lines) in the momentum space, together with the C lines (dashed lines) and the diabolic point (DP). Away from the ideal conditions the DP would reduce into exceptional points (EP) (b) Illustration of polarizations projected on Poinare sphere on the circular path around DP for different Φ. As Φ evolve from -0.25 to 0.7, the path must sweep throught the north/south pole of the Poinare sphere.

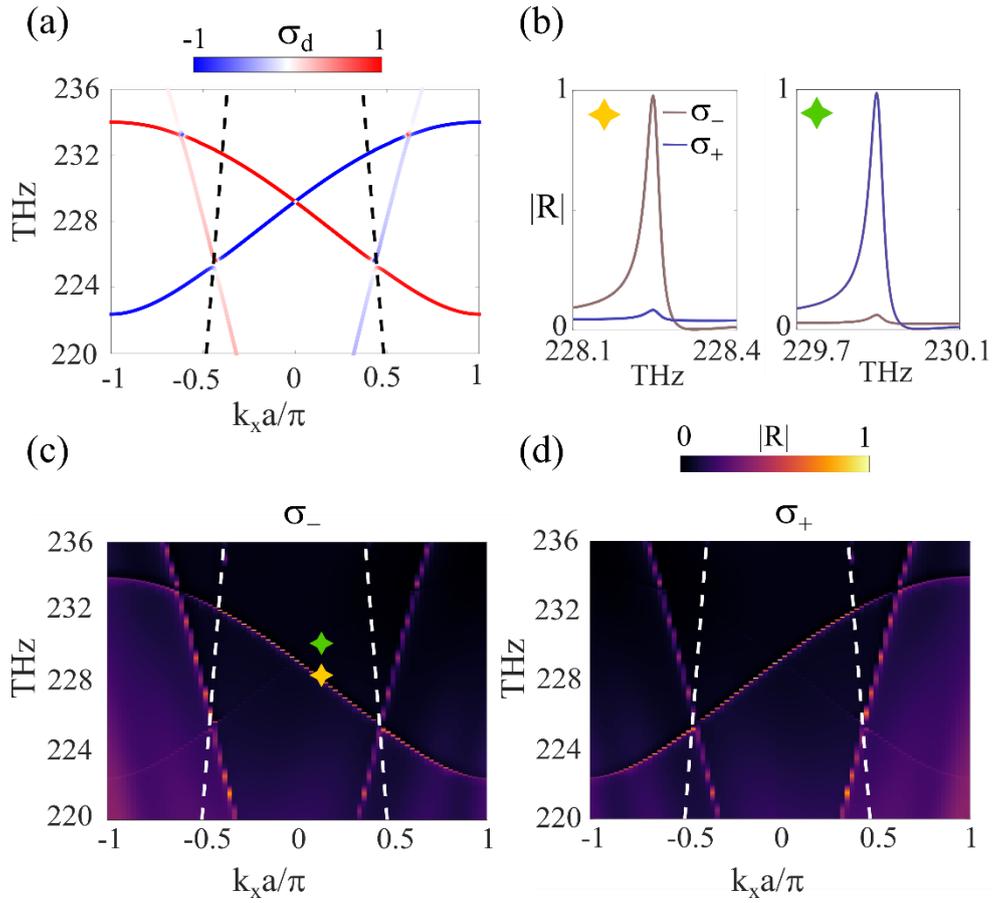

Figure. 4 (a) Dispersions and $\sigma_d$ parameters of the metasurface's eigenmodes along $k_x$ at the Diabolic point with $k_y = 0.172\pi/b$. Light line is depicted as the black dashed lines. An excessive mode exists to interfere with the circularly polarized eigenstates. (b) Detailed spectrum near to the eigen frequencies indicated as the stars in (c) at $k_x = 0.1\pi/a$. The non-vanishing resonance peak in the low reflectivity band and the less than unity (c-d) two dimensional spectrum plot of the reflectivity under left handed circular polarization ($\sigma_-$) and right handed circular polarization ($\sigma_+$) incident.

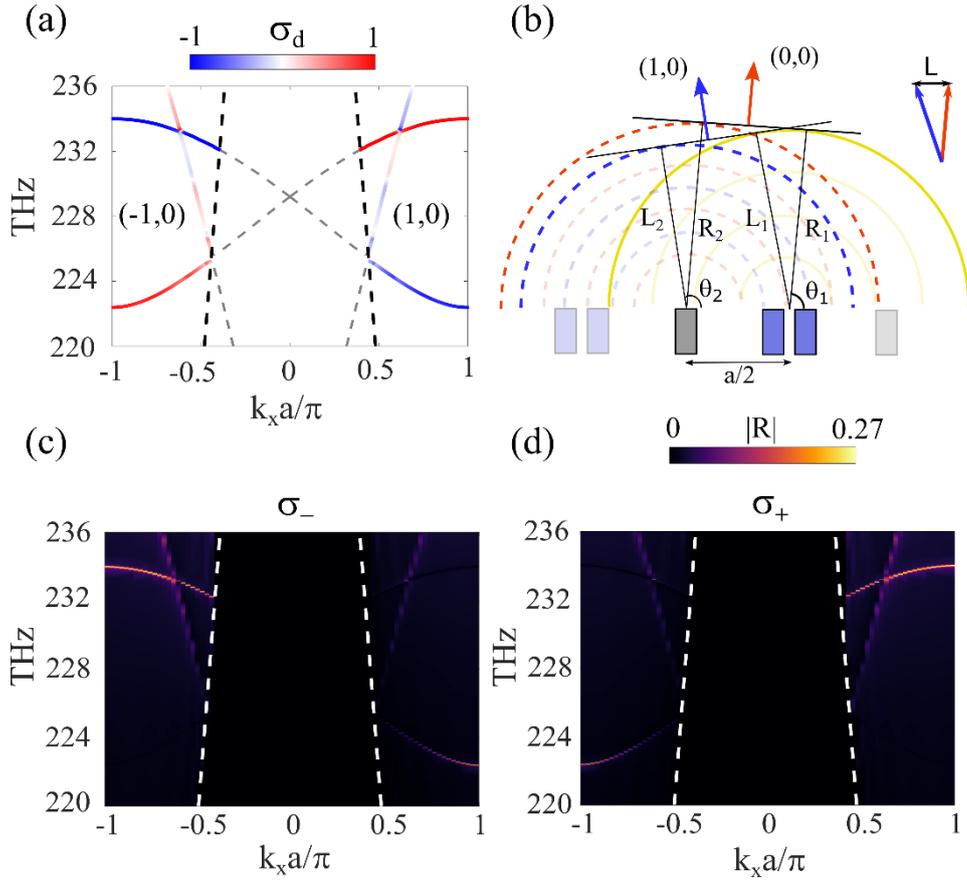

Figure. 5 (a) Dispersions and the ($\pm 1,0$) diffraction orders' $\sigma_d$ parameters of the metasurface's eigenmodes along $k_x$ at the Diabolic point with $k_y = 0.172\pi/b$. Light line is depicted as the black dashed lines. The handedness of the circular polarizations are opposite to the $0^{th}$ order. (b) A geometry explanation of the handedness flip from $0^{th}$ to the ($\pm 1,0$) diffraction orders (c) at $k_x = 0.1\pi/a$. The nonvanishing resonance peak in the low reflectivity and the less than unity (c-d) two dimensional spectrum plot of the diffraction efficiency into diffraction orders under left circular polarization ($\sigma_-$) and right handed circular polarization ($\sigma_+$) incident.